\documentclass[%
prl,
 superscriptaddress,
 amsmath,amssymb,
 reprint,%
]{revtex4-2}

\usepackage{mathbbol}
\usepackage{graphicx}
\usepackage{dcolumn,booktabs}
\usepackage[
breaklinks,
colorlinks,
linkcolor=blue,
citecolor=blue,
urlcolor=blue]{hyperref}

\usepackage{color,soul}

\begin{document}

\title{Convolutional neural networks for long-time dissipative quantum dynamics}

\author{Luis E. Herrera Rodr\'iguez}
\affiliation{Departamento de F\'{i}sica, Universidad Nacional de Colombia, Carrera 30 No. 45-03, Bogot\'{a} D.C., Colombia}
\affiliation{Escuela de Ciencias B\'asicas, Tecnolog\'ia e Ingenier\'ia, Universidad Nacional Abierta y a Distancia, Facatativ\'a, Colombia}
\affiliation{Department of Physics and Astronomy, University of Delaware, Newark, DE 19711, USA}
\author{Alexei A. Kananenka}
\affiliation{Department of Physics and Astronomy, University of Delaware, Newark, DE 19711, USA}
\email{akanane@udel.edu}

\date{\today}

\begin{abstract}
Exact numerical simulations of dynamics of open quantum systems often require immense computational resources.
We demonstrate that a deep artificial neural network comprised of convolutional layers is a powerful tool for
predicting long-time dynamics of an open quantum system provided the preceding short-time dynamics of the system is known. 
The neural network model developed in this work simulates long-time dynamics efficiently and very accurately across different dynamical
regimes from weakly damped coherent motion to incoherent relaxation. The model was trained on a data set relevant 
to photosynthetic excitation energy transfer and can be deployed to study long-lasting quantum coherence phenomena 
observed in light-harvesting complexes. Furthermore, our model performs well for the initial conditions different 
than those used in the training. Our approach considerably reduces the required computational resources for long-time simulations 
and holds promise for becoming a valuable tool in the study of open quantum systems.

\end{abstract}

\maketitle

%%%%%%%%%%%%%%%%%%%%%%%%%%%%%%%%%%%%%%%%%%%%%%%%%%%%%%%%%%%%%%%%%%%%
%
% Introduction
%
%%%%%%%%%%%%%%%%%%%%%%%%%%%%%%%%%%%%%%%%%%%%%%%%%%%%%%%%%%%%%%%%%%%%
\section{Introduction}
Quantum systems encountered in the real world are never completely isolated from their environment. 
It is the interaction between an open quantum system and its environment that alters the otherwise unitary  
dynamics of the system and causes energy dissipation and the destruction of phase coherence~\cite{breuer02,weiss12,leggett87}.
Understanding the temporal evolution of open quantum systems 
is a key problem of a broad interest in chemical physics, quantum optics, quantum biology, ultrafast spectroscopy,
quantum computing, and quantum information technology~\cite{may11,nitzan13,valkunas13,nielsen11,vega17,mukamel95}.

Dynamics of an open system, induced by the Hamiltonian evolution of a total system,
is referred to as reduced (system) dynamics and is described by the reduced density operator.
Many numerically exact methods have been developed to simulate
reduced dynamics~\cite{vega17} including the hierarchy of equations of motion (HEOM)
technique~\cite{tanimura89a,ishizaki09c,tanimura20}, multi-configurational time-dependent Hartree (MCTDH)~\cite{meyer90},
stochastic Liouville--von Neumann equation~\cite{stockburger02}, time evolving density matrix using 
orthogonal polynomials algorithm (TEDOPA)~\cite{prior10}, 
diagrammatic quantum Monte Carlo~\cite{cohen11,cohen15,hsingta17}, path-integral Monte Carlo~\cite{kast13},
and the quasi-adiabatic propagator path-integral (QUAPI) approach~\cite{makarov94,makri92,makri95,makri98}.
Unfortunately, many of these methods %are computationally too expensive and typically 
require computational resources that scale exponentially with the number of simulated time steps and the size of the system
 rendering such methods not efficient or, even completely impractical, to study long-time quantum dynamical phenomena.

One example of such phenomena is furnished by nature.
Photosynthesis is the natural process that provides the energy source for nearly all life on Earth.
Photosynthetic conversion of  energy starts with the absorption of a photon of sunlight by a
light-harvesting pigment and is followed by the excitation energy transfer (EET) to the reaction center.
Recent experiments showed that quantum coherence between electronic states of light-harvesting
complexes can persist for several hundreds 
of femtoseconds even at physiological temperature~\cite{lee07c,engel07,collini10,panitchayangkoon10,harel12,ishizaki12}.
The physical mechanisms underlying the long-lasting quantum coherence and the role of protein environment in EET 
 are still subjects of active investigations. 

Information about the underlying dynamical correlations in open quantum systems is encoded at the initial stages of their 
evolution. Therefore, it is possible to obtain long-time dynamics of open quantum systems from the knowledge of their short-time evolution
bypassing the need for expensive direct long-time simulations. 
For example, the Nakajima--Zwanzig generalized quantum master equation (GQME)~\cite{nakajima58,zwanzig60b} furnishes a general
and formally exact prescription for achieving this goal provided the memory kernel is known~\cite{shi03f,kelly13}. 
Unfortunately, it is difficult to directly calculate the memory kernel exactly for arbitrary systems and solve the GQME.
The latter issue is resolved in the transfer tensor method (TTM)~\cite{cerrillo14,kananenka16,buser17,gelzinis17}.
%whose central objects---transfer tensors are straightforwardly related to the memory kernels. Transfer tensors
%can be easily obtained from a set of dynamical maps which, in turn, are generated over some short period 
%of time by an external numerical technique (e.g. HEOM). The propagation of 
The TTM, however, requires a set of dynamical maps to be supplied by an external numerical technique. %(e.g. HEOM). 
%Then the transfer tensors can be used to propagate system density operator to arbitrary long times
%with the computational cost similar to solving a time-convolution master equation.
%to provide
%subsystem RDM to arbitrary long times amounts to repetitive tensor multiplications. 
For a $N$-level system an input set of dynamical maps generated for the
$N^2$ linearly independent initial conditions of the system is required. Since this input must be provided by some, preferably exact,
method the computational cost of generating such an input grows steeply with the system size.
%TTM is similar to QUAPI approach~\cite{makarov94}. %,makri95a,makri95b}.

In the last decade, the field of deep learning has enjoyed a dramatic expansion owing to multiple
successful applications in the recognition of complex patterns~\cite{lecun15}. Learning temporal and spatio-temporal 
information has been an active area of research recently due to a myriad of potential applications
including detection and characterization of behavior from video
sequences~\cite{dollar05,weinland07,gorelick07,kim07a,klaeser08,wang09a,jia08,liu08b,laptev08,liu09c,taylor10,baccouche10,niebles10,baccouche11,ji13,simonyan14,tran14,karpathy14,sun15,ng15,courtney20}, natural language processing~\cite{hochreiter97,kim14}, and
weather forecasting~\cite{shi15,chen20}.
Recurrent neural network (RNN) is a class of artificial neural networks (ANN) developed specifically
 to analyze temporal data. RNNs take into account the sequential context using recurrent connections in the 
 hidden layers. RNNs based on Long Short-Term Memory (LSTM)~\cite{hochreiter97} cells
outperform other RNNs and found numerous applications~\cite{gers03,baccouche10,ng15}.
%LSTMs maintain an internal state with an additional self connection to pass the information along every time step.

Three-dimensional (3D) convolutional neural networks (CNN)~\cite{ji13} are
based on a straightforward extension of the established 2D CNNs~\cite{lecun98,lecun10} to the 3D spatio-temporal domain.
%3D CNNs learn convolution kernels in both space and time. 
3D CNNs have also been successful in analyzing spatio-temporal data~\cite{ji13}
and are known to outperform LSTMs in certain tasks~\cite{tran14}. 
LSTMs process elements one at a time with a form of memory and thus perceived to be more suitable for extracting 
long-range temporal dependencies
as opposed to convolutions that operate on local neighborhoods. On the other hand LSTMs are more difficult 
to train than CNNs~\cite{courtney20}.

%The dramatic development of machine-learning techniques has 

Despite widespread popularity in the domain of  the physical sciences~\cite{carleo19} machine learning applications to 
open quantum systems have only recently begun to emerge~\cite{schuld19}. 
Restricted Boltzmann machine representations of the density matrix
were used to determine steady states~\cite{vicentini19,nagy19,hartmann19,yoshioka19} and 
Markovian dynamics~\cite{hartmann19} of open quantum systems.
Deep RNNs were used to simulate the dynamics of the spin-boson and Landau--Zener models~\cite{bandyopadhyay18,yang20}, 
as well as to learn Lindblad operators of Master equations and predict the future
time evolution of a system~\cite{banchi18}. 

Given the high computational cost of numerically exact simulations 
and the ability of machine learning techniques to learn complex physical phenomena it is highly
desirable to develop a machine learning approach that can accurately predict long-time dynamics of open quantum systems
and, at least partially, eliminate the need for expensive direct calculations. The purpose of this article is to introduce such a method.
%Accurate prediction of long-time evolution  with 
The physics of dissipative open quantum systems is incredibly rich~\cite{breuer02,weiss12,leggett87}
 naturally rendering such learning problem very difficult.
We illustrate that a properly trained ANN can very accurately predict long-time dynamics of open quantum systems. In contrast to existing
methods~\cite{banchi18,bandyopadhyay18} our approach employs a simpler and easier to train convolutional ANN and,
most importantly, the single ANN developed in this work accurately predicts non-Markovian reduced system dynamics 
in a broad range of dynamical regimes from weakly damped coherent motion to the incoherent decay.

%%%%%%%%%%%%%%%%%%%%%%%%%%%%%%%%%%%%%%%%%%%%%%%%%%%%%%%%%%
%
% Methods
%
%%%%%%%%%%%%%%%%%%%%%%%%%%%%%%%%%%%%%%%%%%%%%%%%%%%%%%%%%%

\section{Theoretical background}
We consider the common system-bath model with the Hamiltonian given by
\begin{equation}
\hat{H} = \hat{H}_s + \hat{H}_{sb} + \hat{H}_b, \label{eq:h}
\end{equation}
where $\hat{H}_s$ is the system Hamiltonian, 
$\hat{H}_b$ describes the thermodynamic reservoir or heat bath, and $\hat{H}_{sb}$ is the coupling between the
system and the bath.

The system Hamiltonian takes the form
\begin{equation}
\hat{H}_s = \sum_{n=1}^{N} \epsilon_n |n\rangle \langle n| + \sum_{k\neq n =1}^{N} J_{nk}|k\rangle \langle n|, \label{eq:hs}
\end{equation}
where the energy of a state $n$ in the absence of the bath is
denoted by $\epsilon_n$ and $J_{nk}$ is the coupling between the $n$th and the $k$th states.

The bath is responsible for the system's energy fluctuation and dissipation.
It is modeled by a set of quantum harmonic oscillators 
\begin{equation}
\hat{H}_b = \sum_{j=1}^{N_b} \left( \frac{\hat{p}_{j}^2}{2m_{j}} + \frac{1}{2}m_{j}\omega_{j}^2 \hat{x}_{j}^2\right),
\end{equation}
where the momentum, position, mass, and frequency of the $j$th harmonic
oscillator are given by $\hat{p}_{j}$, $\hat{x}_{j}$, $m_{j}$, and $\omega_{j}$, respectively. 
It should be noted that the harmonic representation of the bath does not imply that
the microscopic potentials of the environment are assumed to be harmonic.
Rather the rationale behind this choice of the representation of the bath is that the reservoir 
with many macroscopic degrees of freedom carry Gaussian fluctuation properties.

The system-bath coupling is taken to be linear in bath coordinates
\begin{equation}
\hat{H}_{sb} = \sum_{n=1}^{N}\sum_{j=1}^{N_b} c_{j}\hat{x}_{j} |n\rangle \langle n|. \label{eq:hsb}
\end{equation}
%where $c_{nj}$ are the coupling coefficients of $j$th bath mode to $n$th state of the system. 
The coupling constants $c_{j}$ can be characterized by a single function of frequency known as the spectral density
\begin{equation}
C(\omega) = \frac{\pi}{2} \sum_{j=1}^{N_b} \frac{c_{j}^2}{m_{j}\omega_j}\delta(\omega - \omega_j).
\end{equation}

Time-evolution of an open quantum system is described by its reduced density operator $\hat{\rho}_s$ 
which is defined as the partial trace of the total density operator $\hat{\rho}$ over the bath degrees of freedom
\begin{equation}
\hat{\rho}_s(t) = \text{Tr}_b \left[ \hat{\rho}(t)\right]. \label{eq:lvn}
\end{equation}
The reduced density operator evolves in time according to
\begin{equation}
\hat{\rho}_s^I(t) = \hat{U}^I(t)\hat{\rho}_s^I(0), \label{eq:rho}
\end{equation}
where the superscript ``I'' indicates the interaction picture with respect to $\hat{H}_s + \hat{H}_b$ in the Liouville space 
which is defined for any Schr\"odinger picture operator $\hat{O}_S$ as $\hat{O}_I(t)\equiv e^{i\mathcal{L}_st}\hat{O}_S$
with $\mathcal{L}_s$ being the system Liouvillian.
The reduced propagator $\hat{U}^I(t)$ is given by
\begin{equation}
\hat{U}^I(t) \equiv \bigg \langle \mathcal{T}_+ \exp \left[ -\frac{i}{\hbar}\int_0^t d\tau \mathcal{L}^I_{sb}(\tau)\right]\bigg\rangle, \label{eq:u}
\end{equation}
where $\langle \ldots \rangle=\mathrm{Tr}_b\left[\ldots e^{-\beta \hat{H}_b}\right]/Z$
is a thermal quantum-mechanical average over bath modes,
%\str{$\text{Tr}_b\left[\ldots \exp(-\beta \hat{H}_b)\right]/Z$}, 
$\beta \equiv (k_BT)^{-1}$ is the inverse temperature,
$Z\equiv\text{Tr}_b\left[e^{-\beta\hat{H}_b}\right]$ is the partition function, and
$\mathcal{T}_+$ orders the $\mathcal{L}^I_{sb}(t)$'s in increasing time from right to left.

In what follows we show that given the complete knowledge of the short-time evolution of the reduced
density matrix (RDM)
 at the discretized times $\{\rho^s_{ij}(t_k), \rho^s_{ij}(t_{k+1}), \ldots, \rho^s_{ij}(t_m), \forall i,j \in[1,N]\}$, where 
$\rho^s_{ij}(t_k)=\langle i |\hat{\rho}_s(t_k)| j\rangle$, $t_k=k\Delta t$  
and $\Delta t$ is the time step,   
a trained ANN is capable of propagating the RDM to arbitrary long times
$\rho^s_{ij}(t_m) \to \rho^s_{ij}(t_{m+1}), \ldots, \rho^s_{ij}(\infty)$. Thus our approach completely avoids a computationally expensive 
numerically exact propagation of the reduced density operator 
beyond some finite time $t_\mathrm{mem}=t_{m-k}$ which will be referred to as memory or learning time.
The finite memory time is not an assumption but, rather,
a direct consequence of the finite time span of bath correlations in realistic systems.

\section{Methods}
\subsection{Model system}
Even though the approach described in this work is general, 
we present it on the example of a two-level system ($N=2$, $|n\rangle\in\{|0\rangle,|1\rangle\}$) 
coupled to harmonic bath. %comprising $N_b=100$ modes. 
The spectral density is taken to be the Drude--Lorenz density~\cite{mukamel95} (an Ohmic distribution with a Lorentzian cutoff)
\begin{equation}
C(\omega) = 2\lambda \frac{\omega \omega_c}{\omega^2 + \omega_c^2},
\end{equation}
where
$\lambda$ is the reorganization energy which represents the magnitude of fluctuations and dissipation,
and $\omega_c$ is the cut-off frequency which
characterizes how quickly the bath relaxes toward equilibrium. %The latter is related to the bath correlation time by $\tau_c = \omega_c^{-1}$.
 
 \begin{figure*}
\centering
\includegraphics[width=0.99\textwidth]{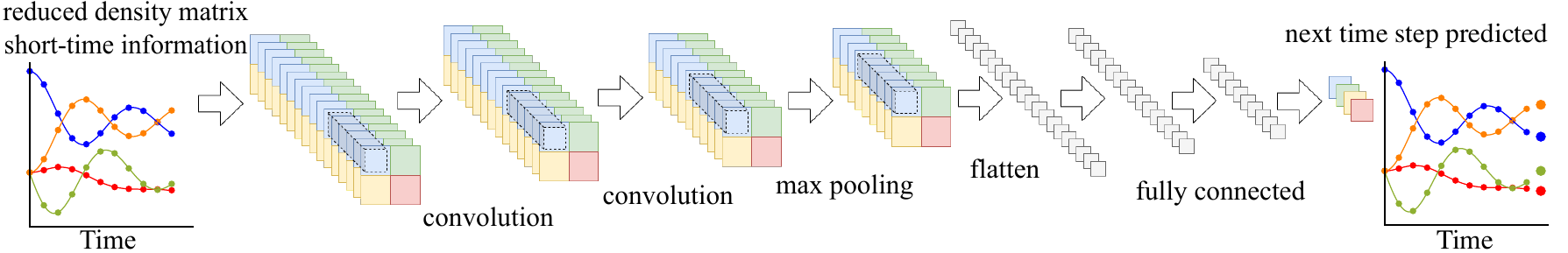}
\caption{Long-time quantum dynamics with artificial neural networks. A set of short-time  
system reduced density matrices at the discretized times serves as an input. It is then processed
through the two convolutional layers followed by the max-pooling layer and two fully-connected layers. The result is the
reduced density matrix predicted for the time step following the last time step of the input set of reduced density matrices.} %See text for
%specific details of the neural network architecture.} 
\label{fig:1}
\end{figure*}

The total system under study is, thus, fully determined
by the five independent energy scales: the energetic bias 
$\Delta=\epsilon_1-\epsilon_0$, the coupling strength $J$,
the reorganization energy $\lambda$, the cut-off frequency $\omega_c$, and thermal energy $k_BT$ of the bath. 
Therefore, in order to create a data set suitable for machine learning framework aimed at predicting quantum dynamics
in all physically realizable regimes, four out of five parameters above have to be extensively sampled. 
While nowadays this task most certainly does not present a major
challenge, in this work we generated only a subset of the total required data set by fixing two
parameters: the coupling strength $J$ and the cut-off frequency $\omega_c$. 

For concreteness we illustrate our approach on the example of photosynthetic EET and consider a
two-molecule system or dimer. The system Hamiltonian, Eq.~\ref{eq:hs}, describes an electronic 
system within a single-excitation manifold %of electronic states 
which for a dimer comprises of singly-excited 
electronic states of each molecule. Thus, in Eqs.~\ref{eq:hs} and~\ref{eq:hsb}
$|n\rangle$ represents the state where only one of the two molecules is in its excited electronic state
while the other molecule is in its ground electronic state. %: $|n\rangle\equiv|\psi_{ne}\rangle \prod_{n\neq k}|\psi_{kg}\rangle$. 
The ground and excited states of the total system are denoted as $|0\rangle$ and $|1\rangle$ respectively. %$|0\rangle=|\psi_{0\mathrm{e}}\rangle |\psi_{1\mathrm{g}}\rangle$ and $|1\rangle=|\psi_{0\mathrm{g}}\rangle |\psi_{1\mathrm{e}}\rangle$,
%where ``0'' and ``1'' label a molecule, and ``g'' and ``e'' refer to the ground and excited state respectively. 
The single-excitation %one-exciton 
manifold approximation is often well-justified. 
For example, in certain photosynthetic organisms, such as purple bacteria, the rate of photon capture is low enough
to guarantee that at most one excitation is present on the photosynthetic membrane~\cite{fassioli09}.

Following previous work (e.g., Refs.~\cite{amerongen00,ishizaki09,ishizaki09c,ishizaki12}) 
we choose $J=100$ cm$^{-1}$ and set the cut-off frequency $\omega_c$ 
to 53 cm$^{-1}$. These are the typical parameters of photosynthetic EET. We note that, according to Ref.~\cite{rebentrost11},
specific values of $J$ and $\omega_c$ chosen here correspond to a moderately non-Markovian regime.

\subsection{Data generation and preprocessing}
Building a representative data set is an important first step in every machine learning project. 
The HEOM method implemented in the \textsc{PHI} code~\cite{strumpfer12a} is used to solve Eqs.~\ref{eq:rho} and~\ref{eq:u}.
The hierarchy truncation is set to 20
and 3 Matsubara terms are used. The integration time step is set to 0.1 fs. The convergence of all the results,
presented in this article, with respect to these parameters
is verified. The data set containing time-evolved RDMs is generated for
all combinations of the following parameters: $\Delta\in\{0,25,50,75,100,125,150,175, 200, 225, 250, 275, 300\}$
cm$^{-1}$, $\lambda \in \{2, 10, 20, 30, 40, 50, 60, 70, 80, 90, 100, 120, 130$, $150, 200, 250, 300, 350, 400, 450, 500\}$ cm$^{-1}$,
and $T\in \{25, 50, 75, 100, 125, 150, 175, 200, 225, 250, 275, 300, 325$, $350, 375, 400\}$ K.
%A total of 4,368 HEOM calculations were performed.
%The data set used in this work is generated by varying temperature from 25 to 400 K
%and the electronic energy gap $\Delta$ from 0 to 300 cm$^{-1}$. 
We note that while the suggested reorganization energies in EET
are typically similar to the electronic coupling~\cite{brixner05,cho05}, %. However, 
for the purpose of building a diverse data set,
the range of reorganization energies is extended to as low as 2 cm$^{-1}$ and
as high as 500 cm$^{-1}$. 
%to the lowest value 
%of 2 cm$^{-1}$ and the highest value of  400 cm$^{-1}$. 
The resulting data set is diverse enough to 
represent weakly damped coherent dynamics $(J/\lambda > 1)$, incoherent  
decay $(J/\lambda < 1)$, and the transition between the two regimes.

For each combination of parameters the reduced density operator is 
propagated %according to Eqs.~\ref{eq:rho}-\ref{eq:u} 
starting from the factorized product state 
$\hat{\rho}(0)=\hat{\rho}_s(0)\otimes\hat{\rho}_b^{eq}$. 
The electronic system is assumed to start out at the first excited state $\hat{\rho}_s(0)=|1\rangle\langle1|$,
and the bath is at thermal equilibrium
$\hat{\rho}_b^{eq}=e^{-\beta \hat{H}_b}/Z$. %, where $Z=\text{Tr}[ e^{-\beta \hat{H}_b} ]$ and $\beta$ is the inverse temperature. 
It should be noted that, in general, this factorized initial state is unphysical
due to the neglect of an inherent correlation between the system and the bath. However, in modeling electronic excitation processes
such initial state is appropriate because it corresponds to the electronic ground or excited state
generated by photoexcitation in accordance to the vertical Frank--Condon transition. 

The total propagation time is set to 1.0 ps for all parameter sets even though in some cases the coherent
dynamics persists for longer times. It is though still longer than a typical coherence time 
of several hundreds femtoseconds found in photosynthetic EET~\cite{ishizaki09c}.

%The resulting quantum trajectory of the RDM is divided into shorter trajectories with the length equal to
%the memory time. Each of these short RDM trajectories is used as an input to an ANN. 
%

 For each set of parameters the generated quantum RDM 
trajectory is divided into shorter RDM trajectories with the length equal to the memory time.
Since the total propagation time is fixed to 1.0 ps, the actual size of the data set used  
varies depending on the length of the memory time. 
For example, for the memory time of 0.2 ps the data set is comprised of 21,840 0.2 ps long RDM trajectories. 
These trajectories are time discretized with the time step of 5 fs.
A small set of 150 RDM trajectories is randomly taken from the total data set and is used
to produce the results presented in this article.
The remaining data set is partitioned into a training set of 80\% of the
data and a validation set of 10\% of the data. Additionally,
10\% of the data is held out during the training procedure
and is used for testing.

\subsection{Neural network model and training}
The ANN architecture used in this work is based on three-dimensional (3D) convolutional and 3D pooling layers.
It is schematically illustrated in Figure~\ref{fig:1}.
3D convolutional layers are essentially 2D convolutional layers extended such that all filters operate over space and time.
Similarly, 3D pooling is a straightforward extension of 2D pooling to the temporal domain. 
3D convolutional layers extract short-time correlations, while pooling layers combine information 
across longer time periods. 

The ANN architecture is optimized using the Hyperopt library~\cite{bergstra13}.
The number of 3D convolutional layers is chosen between 1 and 3.
The size of a 3D convolutional kernel is represented by a triple ($d_t$,$d_{s1}$,$d_{s2}$), where $d_t$ is the temporal depth and $(d_{s1},d_{s2})$
denotes the spatial size. We set the spatial size to be $d_{s1}=d_{s2}=1$ for the reasons explained below. 
The temporal depths of the filters for each layer are chosen as $d_t=2^n$ with all integer values of $n\in[4,8]$ tested.
%The 3D kernels have the shape of $\{x,1,1\}$, where all integer values of $x\in[1,10]$ are tested.
The single 3D maximum pooling layer with the kernel size of (2,1,1) follows 3D convolutional layers. 
The number of fully-connected layers is chosen between 1 and 3 and the number of 
neurons in each fully-connected layer is chosen as 2$^n$ with all integer values of $n \in [4,7]$ tested.
Commonly used activation functions including hyperbolic tangent, 
sigmoid, and the rectified linear activation function (ReLU)~\cite{maas13} are
tested. For the output nodes the linear function $f(x) = x$ is used to eliminate any restriction on the range of possible output values. 
The output layer contains the three real-valued numbers that uniquely define the next time-step RDM:
one diagonal element $\rho_{00}^s(t)$, as well as the real and imaginary parts of one off-diagonal element
$\mathrm{Re}\{\rho_{01}^s(t)\}$, and $\mathrm{Im}\{\rho_{01}^s(t)\}$. The other diagonal element $\rho_{11}^s(t)$
is also added to the output layer in order to check whether our ANN model can learn some
fundamental properties of the density matrix e.g., the unit trace property. %, $\rho_{00}^s+\rho_{11}^s=1$ .

The optimized ANN comprises the two 3D convolutional layers followed by one 3D maximum pooling layer, two fully-connected layers and
the output layer. The optimal number of filters in both 3D convolutional layers is 30.
The temporal depths of the kernels in the first and second 3D convolutional layers are 16 and 4 time 
steps, or 80 and 20 fs correspondingly. 
The two fully-connected layers containing 256 and 64 neurons follow the 3D pooling layer. 
The ReLU activation function is selected for each of the 3D convolutional and fully-connected layers.
The total number of trained parameters in the optimized ANN is 2,816,624.

%While the usfullness of 1x1 convolution may 
%raise questions we note that they have proven effective in various ML tools~\cite{} 
%However, the correlation between them is taken into account by means of 3D pooling layer which operates on the
%output of the second 3D convolutional layer.
%It applies maximum pooling to the stacked data within a 3D pooling cube.

The 1x1 spatial convolution employed in all 3D convolutional layers 
means that the correlation between individual elements of the RDM at a given time are not accounted for. 
It should be noted that in some cases only a limited knowledge of short-time RDMs is available.
This partially motivates our choice to separately process different elements of the RDM through the 3D convolutional and pooling layers. 
Further motivation stems from the observation that increasing $d_{s1}$ and/or $d_{s2}$ to 2 in 3D convolutional and pooling layers
does not increase the prediction accuracy. %lead to any improvement of the ANN performance.

\begin{figure*}
\centering
\includegraphics{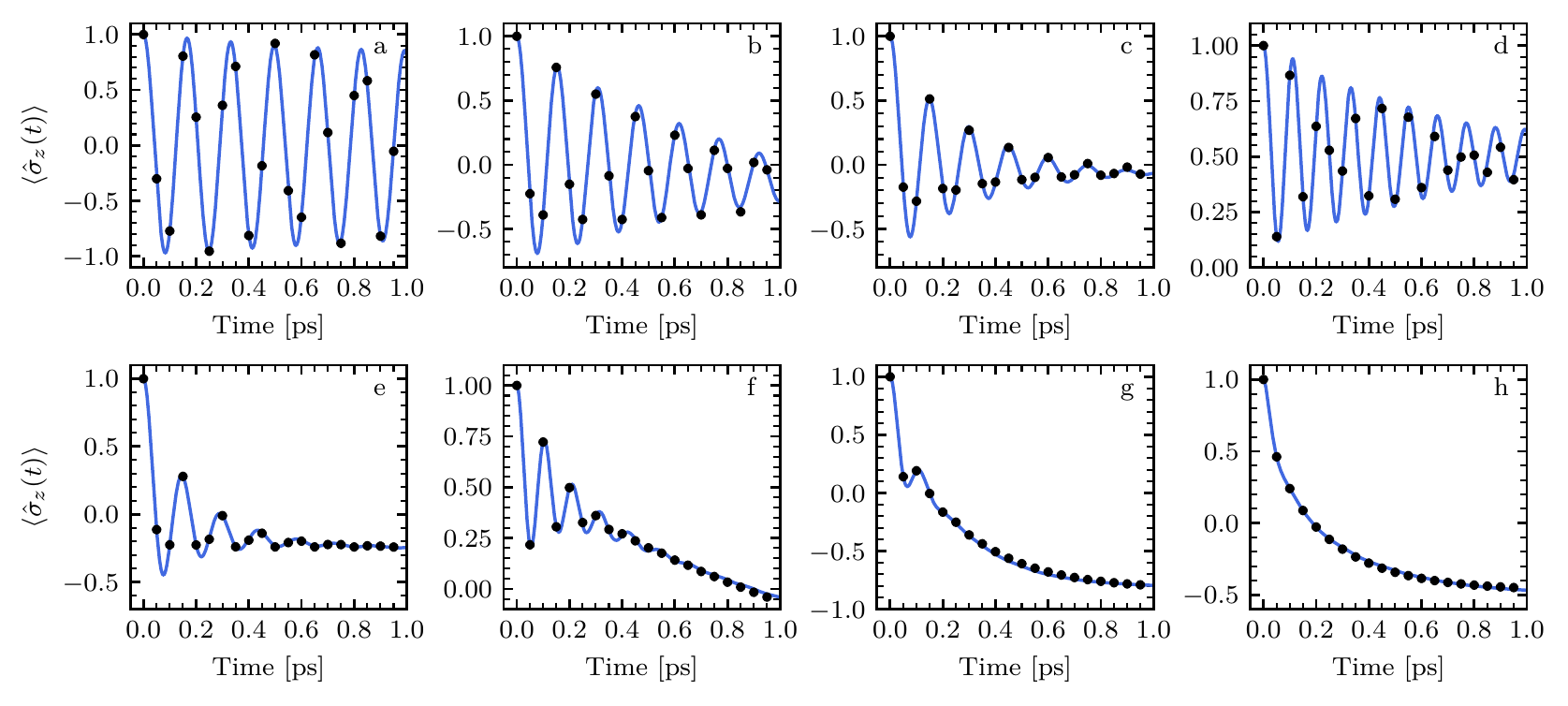}
\caption{Expectation value of $\hat{\sigma}_z$ as a function of time predicted by the neural network model 
developed in this work (blue solid lines) for various sets of parameters compared to the exact results (black circles)
obtained with the HEOM method:
(a) $\Delta$=0 cm$^{-1}$, $\lambda$=2 cm$^{-1}$, $T$=125 K; 
(b) $\Delta$=75 cm$^{-1}$, $\lambda$=10 cm$^{-1}$, $T$=100 K;
(c) $\Delta$=25 cm$^{-1}$, $\lambda$=20 cm$^{-1}$, $T$=275 K;
(d) $\Delta$=225 cm$^{-1}$, $\lambda$=2 cm$^{-1}$, $T$=175 K;
(e) $\Delta$=50.0 cm$^{-1}$, $\lambda$=70.0 cm$^{-1}$, $T$=100 K;
(f) $\Delta$=250 cm$^{-1}$, $\lambda$=20 cm$^{-1}$, $T$=75 K;
(g) $\Delta$=225 cm$^{-1}$, $\lambda$=80 cm$^{-1}$, $T$=50 K;
(h) $\Delta$=175 cm$^{-1}$, $\lambda$=200 cm$^{-1}$, $T$=25 K.
Other parameters are $J=100$ cm$^{-1}$ and $\omega_c=53$ cm$^{-1}$.} 
\label{fig:2}
\end{figure*}

The backpropagation is used to train the ANN to predict the RDM for the time step immediately 
following the chronologically last time step of the input set of RDMs.
In the backpropagation the
adaptive moment solver~\cite{kingma14} is used
and the mean squared error function is employed as the loss function. %in the minimization. %with equal weights for each element of the RDM.
ANN models are trained using the Keras software package~\cite{keras} with
the TensorFlow~\cite{tensorflow} backend.  Further hyperparameter tuning is performed by means of a random-search algorithm~\cite{bergstra12}.
The best results were obtained with the learning rate of 1$\cdot$10$^{-7}$ and a batch size of 512.

%%%%%%%%%%%%%%%%%%%%%%%%%%%%%%%%%%%%%%%%%%%%%%%%%%%%%%%%%%
%
% Results
%
%
%%%%%%%%%%%%%%%%%%%%%%%%%%%%%%%%%%%%%%%%%%%%%%%%%%%%%%%%%
\section{\label{sec:results} Results and discussion}
The ANN introduced above is used to simulate the relaxation dynamics of the dimer
coupled to the harmonic bath described in the previous section. We consider the expectation value of the $\hat{\sigma}_z$ Pauli operator
$\langle \hat{\sigma}_z (t)\rangle = p_1(t) - p_0(t)$, where $p_{1(0)}(t)$ is the time-dependent population of the corresponding state. 
The results for the eight sets of parameters representing different regimes ranging from weakly damped coherent oscillations to
 incoherent relaxation are illustrated in Figure~\ref{fig:2}. In each case the memory time of 0.2 ps is used and the corresponding set of RDMs
is supplied to the ANN as an input. The ANN predicts the RDM at the next time step. Then, a new input is formed by combining the 
newly predicted RDM with all the RDMs taken within the 0.2 ps time span back starting from the just predicted RDM. 
This process is repeated until the
specified propagation end time is reached. We emphasize that only the short-time initial dynamical information is required. The rest of
the time-evolution is simulated by reconstructing the RDM based on ANN predictions for each time step beyond the
initial learning time. Thus, if a single-step 
prediction error is not sufficiently small then the error will very quickly accumulate resulting in a rapid deterioration 
of the accuracy. However, as evident from Figure~\ref{fig:2} our method reproduces long-time dynamics nearly exactly
and irrespective of the dynamical regime. The data shown in Figure~\ref{fig:2} represents a small fraction of a test set.
The interested reader is referred to Supporting Information Figures S1-S5 where many more results are presented. 
In addition to $\langle\hat{\sigma}_z(t)\rangle$, Figures S1-S5 also report the ANN 
predictions for the real and imaginary parts of the off-diagonal elements of the RDM. 
%$\text{Re}\{\rho_{01}(t)\}$ and  $\text{Im}\{\rho_{01}(t)\}$.
These results demonstrate that our ANN model can very accurately predict long-time evolution of the entire RDM. 
%These results further reaffirm the near exact accuracy of our NN model.
Additionally, we note that by virtue of its ability to accurately predict dynamics to arbitrary times, 
our ANN model also correctly reproduces equilibrium populations.

The accuracy of our method depends on the memory time as can be expected. 
Supporting Information Figure S6 illustrates that the ANN prediction error decreases monotonically with increasing memory time.
We choose a memory time of 0.2 ps in all calculations reported in this article 
as it provides a compromise between accuracy and computational cost required
to generate the input set of RDMs. 
%If dynamics is predominantly Markovian then the chosen memory time can be 
%longer than the required memory time and, conversely, for a pronounced non-Markovian regimes it might be too short.
More rigorously, the learning time can be determined from the decay of the magnitude of 
transfer tensors, as it is done in the TTM method~\cite{cerrillo14,gelzinis17}, or a norm of the GQME memory kernel.
A pre-defined and identical memory time for all cases studied here is chosen to accommodate the fixed-size 
input requirement of ANNs. %Furthermore, we aim to develop a general
%ANN model capable of reconstructing long-time dynamics in all ranges of parameter space.
It is highly desirable to provide a further flexibility to our approach by designing a descriptor that can 
integrate various memory times and be a suitable input to ANN. The work in this direction is underway.

\begin{figure}
\centering
\includegraphics{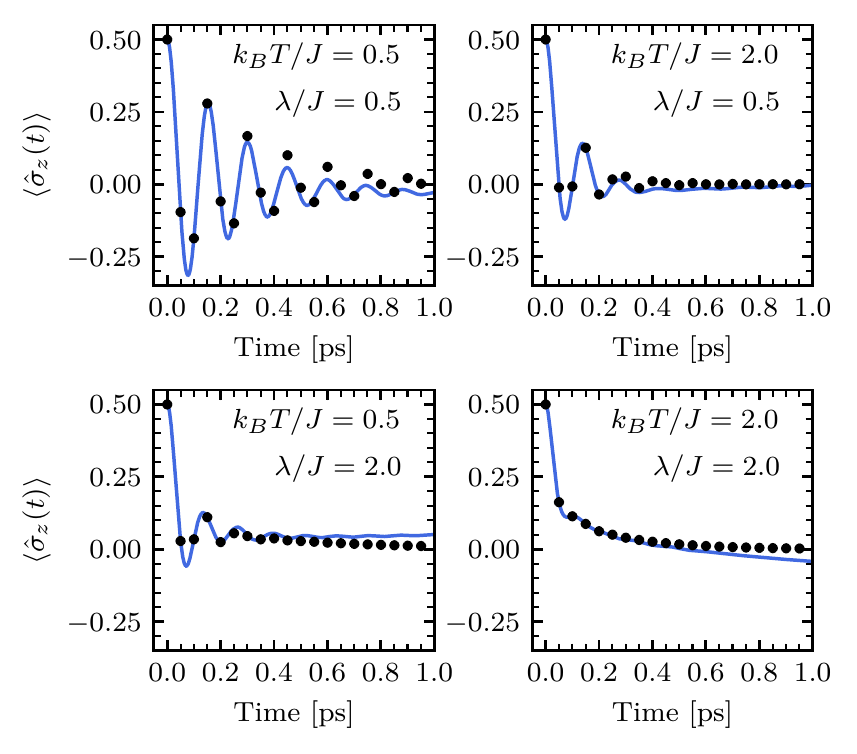}
\caption{Expectation value of $\hat{\sigma}_z$ as a function of time predicted by the neural network model developed in this work
(blue lines) compared to the exact result (black lines) for the initial system reduced density matrix $\rho_s(0) = (1/4) |0\rangle \langle0 | + (3/4) |1\rangle \langle1 |$,
$J$=100 cm$^{-1}$, $\Delta=0$ cm$^{-1}$, and $\omega_c=53$ cm$^{-1}$.   } 
\label{fig:3}
\end{figure}

As discussed in Ref.~\cite{rebentrost11}, which utilizes the measure of non-Markovianity based on
quantum state trace difference~\cite{breuer02}, for fixed $J$ and $\omega_c$ non-Markovianity is maximized 
for intermediate values of $\lambda\sim$40--80 cm$^{-1}$
and site energy differences $\Delta$ below $\sim$150 cm$^{-1}$. Additionally, we note that
the typical values of $\omega_c$ for photosynthetic EET satisfy the high-temperature assumption $\beta\hbar\omega_c<1$~\cite{ishizaki09c},
which for $\omega_c=53$ cm$^{-1}$ corresponds to temperature $T>77$ K.  All of these conditions are satisfied in Figure~\ref{fig:2}e,
which thus provides a test of our ANN model in a realistic scenario of
considerable non-Markovian dynamics. We observed a very good agreement between the 
ANN predicted long-time dynamics and the exact results in this case as well.

The ability of our ANN model to learn the fundamental properties of the density matrix
is tested on the unit trace and positive semidefiniteness properies. %of the RDM 
We emphasize that these properties are not enforced during the training.
%but rather used as a test of the ability of our ANN model to learn the fundamental properties of the density matrix.
Impressively, we found that our ANN model was able to learn both properties essentially precisely. The largest deviation of the trace of
the predicted RDMs from unity in a held out data set of 150 samples is only 1$\cdot$10$^{-5}$ and not a single
predicted RDM is found to have negative diagonal elements. %violate the positive semidefiniteness property.

As a stringent test of our ANN model we assess its ability to generalize to the input data not present
in the original data set. To this end we evaluate our ANN model on an input set of short-time RDMs corresponding to the
following  initial mixed state of the electronic
system $\hat{\rho}_s(0) = (1/4) |0\rangle \langle0 | + (3/4) |1\rangle \langle1 |$. The bath is still kept at thermal equilibrium. 
Figure~\ref{fig:3} shows the results 
obtained for the four cases ranging from small to large reorganization energy and low to high temperature. 
The electronic system parameters are fixed to be $J=100$ cm$^{-1}$, $\Delta=0$ cm$^{-1}$ and the cut-off frequency 
$\omega_c$ is set to 53 cm$^{-1}$.
As anticipated, our ANN model performs slightly worse in the region of small reorganization energy and low temperature which is
the most difficult regime of weakly damped coherent oscillatory dynamics. 
Overall across all the cases studied with the out-of-data set initial condition of the system we notice that the single time-step 
prediction error slightly increases, compared to the results shown in Figure~\ref{fig:2}, 
which results in noticeable deviations of the predicted populations from the exact ones at long times. 
However, these deviations are still small and do not exceed 3.6\%.
Thus, we conclude that our ANN model performs
reasonably well even for the initial condition of the system different from the initial condition used to generate the original data set.
It should be noted that extending our ANN model to regimes of arbitrary electronic coupling and bath correlation time 
is less straightforward and would require a retraining with a more appropriate data set.

\section{Conclusions}
In summary, we showed that an ANN trained on a set of time-discretized RDMs is capable of predicting the 
future time-evolution of a given RDM with high accuracy. 
Only a short-time calculation with numerically accurate method is required and the rest of the dynamics
can be efficiently predicted by our ANN model. Thus our approach considerably reduces the required resources for
long-time simulations of dissipative open quantum systems.
What makes our approach particularly appealing is (i) the relative simplicity of the ANN model employed
as opposed to ANNs with internal
memory that are more difficult to train and (ii) its ability to accurately
predict long-time dynamics of open quantum systems across various dynamical regimes including nonadiabatic
and non-Markovian dynamics. 
A photosynthetic EET %a few hundreds of femtoseconds long quantum coherences 
was chosen as an example but the same approach can be used to study other phenomena 
provided the relevant data sets are available. 
The total data set used in this work is built from 4,368 HEOM 
calculations. Given the currently available computational
resources the generation of similar data sets to study other problems is a trivial task.
%Furthermore, once generated, such data sets can be reused in subsequent studies. 
Our approach is yet another illustration of the ability of ANNs to model 
complex physical phenomena at a small fraction of the cost of a full numerically exact calculation.

\section{Acknowledgements}
This work was supported by the startup funds of the College of Arts and Sciences and the
Department of Physics and Astronomy of the University of Delaware. 
L. E. H. R. would like to
acknowledge support by the Beyond Research Program between University of 
Delaware and Universidad Nacional de Colombia.
Calculations were performed with high-performance computing
resources provided by the University of Delaware. 

%\end{document}

\end{document}